\documentclass[twocolumn,prl,showpacs,preprintnumbers,superscriptaddress,amsmath,amssymb,citeautoscript]{revtex4-1}
\pdfoutput=1
\usepackage{graphicx}
\usepackage{epstopdf}
\usepackage{dcolumn}
\usepackage{color}
\usepackage{amsmath,amssymb}
\usepackage[english]{babel}
\usepackage{bm}
\usepackage[hyperindex]{hyperref}
\hypersetup{colorlinks,citecolor=blue, filecolor=blue,linkcolor=blue,urlcolor=blue}

\usepackage{multirow}
\usepackage{rotating}
\usepackage{array}

\newcommand{\bpm}{\begin{pmatrix}}
\newcommand{\epm}{\end{pmatrix}}
\newcommand{\bs}{\boldsymbol}
\newcommand{\be}{\begin{equation}}
\newcommand{\ee}{\end{equation}}
\newcommand{\beq}{\begin{eqnarray}}
\newcommand{\eeq}{\end{eqnarray}}

\DeclareMathOperator{\sgn}{sgn}

\DeclareMathOperator{\im}{Im}

\begin{document}

\title{Obtaining Majorana and other boundary modes from the metamorphosis of impurity-induced states: exact solutions via T-matrix}
\author{Vardan Kaladzhyan}
\email{vardan.kaladzhyan@phystech.edu}
\affiliation{Department of Physics, KTH Royal Institute of Technology, Stockholm, SE-106 91 Sweden}
\author{Cristina Bena}
\affiliation{Institut de Physique Th\'eorique, Universit\'e Paris Saclay, CEA
CNRS, Orme des Merisiers, 91190 Gif-sur-Yvette Cedex, France}

\date{\today}

\begin{abstract}

We provide here a new and exact formalism to describe the formation of end, edge or surface states through the evolution of impurity-induced states.  We propose a general algorithm that consists of finding the impurity states via the T-matrix formalism and showing that they evolve into boundary modes when the impurity potential goes to infinity. We apply this technique to obtain Majorana states in 1D and 2D systems described by the Kitaev model with point-like and respectively line-like impurities. We confirm our exact analytical results by a numerical tight-binding approach. We argue that our formalism can be applied to other topological models, as well as to any model exhibiting edge states. 
\end{abstract}

\maketitle

The discovery of quantum physics in the beginning of the twentieth century significantly accelerated  the progress in solid state physics. One of the oldest and challenging problems in this field is taking into account the presence of boundaries. As Pauli once said, "God made the bulk; surfaces were invented by the devil" \cite{Jamtveit1999}.

Various methods were developed to treat boundaries. One of the most common techniques is the numerical diagonalization of a tight-binding Hamiltonian with open boundary conditions \cite{Slater1954,Busch1987}. Analytically, however, the formation of boundary modes is usually studied by solving the Schr\"odinger equation with specific boundary conditions \cite{Davison1992}. The latter is sometimes cumbersome and oftentimes requires making specific approximations which decrease the generality of the obtained wavefunctions of the boundary modes. Less common techniques rely on the use of boundary Green's function \cite{Sancho1984,Umerski1997,Peng2017} and the bulk-boundary correspondence  \cite{Rhim2018}.

Here we propose a completely novel, general and exact technique to obtain the energies and the wavefunctions of boundary modes in systems of arbitrary dimensions. Instead of solving the problem of a finite-size system with a desired boundary, we suggest to consider an infinite system with  a localized impurity which follows the shape of the boundary. We subsequently obtain the corresponding impurity-induced states using the T-matrix formalism \cite{Balatsky2006}.  As intuitively expected, we show that by taking the impurity potential to infinity we recover the formation of end, edge or surface states, depending on the dimension of the system.

For the sake of clarity, we exemplify our proposal by focusing on the formation of Majorana end modes in a Kitaev chain \cite{Kitaev2001} and of Majorana chiral edge states in a 2D system described by the spinless Kitaev model \cite{Volovik1999,Read2000,Ivanov2001}; we show that the analytical T-matrix formalism is entirely consistent with a numerical tight-binding calculation. However, our technique can as well be applied other systems supporting both topological and trivial boundary modes, such as models combining $s$-wave superconductivity, spin-orbit coupling and a Zeeman field \cite{Oreg2010,Lutchyn2010}, topological insulators \cite{Fu2008,Qi2011}, graphene \cite{Yao2009} and Weyl and Dirac materials \cite{Wan2011,Okugawa2014}; we have also checked the validity of our formalism for these systems \cite{unpublished2018}. Our results are in agreement with previous work \cite{Slager2015} proposing impurities
as local probes of topology in band insulators. 
\begin{figure}[t!]
\begin{center}
\vspace{.2in}
\includegraphics[width=0.9\columnwidth]{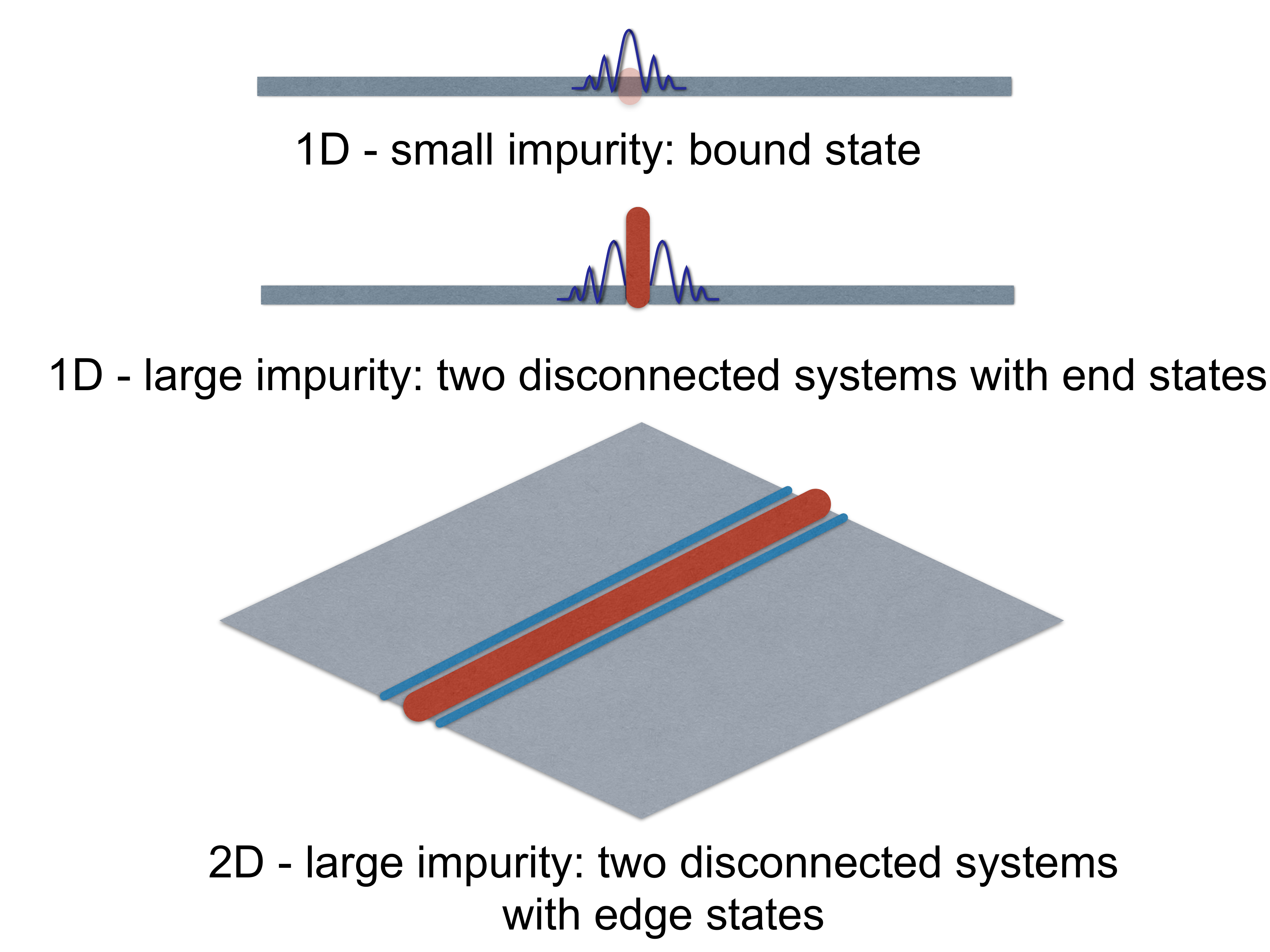}
\end{center}
\caption{A simple exemplification of a 1D system with a localized impurity: when the impurity potential goes to infinity this is equivalent to two disconnected semi-infinite wires. Similarly, a 2D infinite system with a line-like infinite-potential impurity is equivalent to two disconnected half-planes.}  
\label{fig1}
\end{figure}

\vspace{-3mm}
We thus suggest the following four-step algorithm for finding end (edge, surface) states: 
\vspace{-1mm}
\begin{enumerate}
	\setlength\itemsep{-0.2em}
	\item Take an infinite 1D (2D, 3D) system
	\item Introduce a point-like (line-like, plane-like) scalar impurity described by a delta-function potential 
	\item Use the T-matrix formalism to find the energies and wavefunctions of the impurity-bound states 
	\item Formally set the impurity potential to infinity to recover the formation of boundary modes.
\end{enumerate}

\textit{T-matrix formalism.} We start by presenting briefly the theoretical framework required to implement this algorithm. We denote the momentum-space Hamiltonian of a given system $\mathcal{H}_{\bs{p}}$, and we define the unperturbed Matsubara Green's function as follows: $G_0(\bs{p},i\omega_n) \equiv \left(i\omega_n - \mathcal{H}_{\bs{p}} \right)^{-1}$, where $\omega_n$ denote Matsubara frequencies. In the presence of an impurity $V_{imp}(\bs{r})$ the latter is modified to
\begin{eqnarray} 
G(\bs{p}_1,\bs{p}_2,i\omega_n)&=&G_0(\bs{p}_1,i\omega_n)\delta(\bs{p}_1-\bs{p}_2)\\&+&
G_0(\bs{p}_1,i\omega_n)T(\bs{p}_1,\bs{p}_2,i\omega_n)G_0(\bs{p}_2,i\omega_n), \nonumber 
\end{eqnarray}
where the T-matrix $T(\bs{p}_1,\bs{p}_2,i\omega_n)$ describes the cumulated effect of all-order impurity-scattering processes \cite{Mahan2000,Balatsky2006}. For the particular case of a delta-function impurity $V_{imp}(\bs{r}) = V \delta(x)$, the form of the T-matrix in 1D is momentum independent and can be written as \cite{Byers1993,Ziegler1996,Salkola1996,Balatsky2006,Bena2008}:
\begin{eqnarray}
T(p_1, p_2, i\omega_n)&=& [\mathbb{I} -V \cdot \int \frac{d p}{2\pi} G_0(p,i\omega_n)]^{-1} \cdot V
\label{Tmatrix1D}
\end{eqnarray}
while in 2D we have
\begin{eqnarray}
\label{Tmatrix2D} &&T(p_{1x},p_{1y},p_{2x},p_{2y},i\omega_n)= \\&&=\delta(p_{1y}-p_{2y}) [\mathbb{I} - V \cdot \int \frac{d p_x}{2\pi} G_0(p_x,p_{1y},i\omega_n)]^{-1} \cdot V \nonumber
\end{eqnarray}
Note that this is independent of $p_{1x}$ and $p_{2x}$ (due to the fact the impurity is a delta potential in the $x$ direction, and reversely, that it is a delta function in $p_{1y}-p_{2y}$ since the impurity is independent of position in the $y$ direction.

%Eqs.~(\ref{Tmatrix1D}) and (\ref{Tmatrix2D}) contain all the information about the bound states formed in the system. 

%In what follows we use the well-established models for 1D and 2D p-wave superconductors to illustrate the approach we have developed.

In what follows we will use this formalism at zero temperature to calculate the retarded Green's function $\mathcal{G}(\bs{p}_1,\bs{p}_2,E)$ obtained by the analytical continuation of the Matsubara Green's function $G(\bs{p}_1,\bs{p}_2,i\omega_n)$ (i.e. by setting  $i \omega_n \rightarrow E+ i\delta$, with $\delta \to +0$).\\

\begin{figure}[h!]
\begin{center}
\includegraphics[width=0.55\columnwidth]{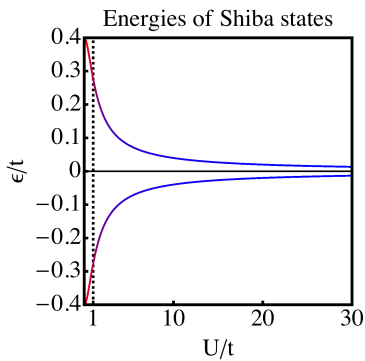}
\end{center}
\caption{The energies of YSR states given by Eq.~(\ref{YSRenergies1D}) plotted as a function of the impurity strength $U$ (in the units of $t$, the hopping parameter). The black dashed line corresponds to $U/t=1$, for which $E/t \approx 0.28$. The energies asymptotically approach zero when $U \to \infty$ marking the metamorphosis of YSR states into Majorana end states. We have set $\mu/t=0$, $\Delta/t=0.4$.}  
\label{fig2}
\end{figure}

\textit{1D Kitaev chain.} We start by considering an infinite spinless Kitaev chain described by the following tight-binding Hamiltonian
\begin{align}
 \mathcal{H}_{TB} = \sum \limits_i -\mu c^\dag_i c^{\phantom{\dag}}_i - \left( t c^\dag_i c^{\phantom{\dag}}_{i+1} - \Delta  c^{\phantom{\dag}}_i c^{\phantom{\dag}}_{i+1} + \mathrm{H.c.} \right)
\label{HKitaevChain}
\end{align}
where $c^\dag_i (c^{\phantom{\dag}}_i)$ are creating (annihilating) operators on the $i$-th site, $t$ is the hopping amplitude, $\mu$ denotes the chemical potential and $\Delta > 0$ is the superconducting pairing amplitude. We set the lattice constant $a$ as well as $\hbar$ to unity. In momentum space the Hamiltonian in Eq.~(\ref{HKitaevChain}) becomes
\begin{align}
\mathcal{H}^{1D}_p =
	\bpm 
		-\mu/2 - t \cos p & i\Delta \sin p  \\
		-i\Delta \sin p & \mu/2 + t \cos p
	\epm.
	\label{Hp1D}
\end{align} 
We introduce a delta-like potential impurity into the chain, localized at $x=0$:
\begin{align}
V_{imp}(x) = U \delta(x)
	\bpm
		1 & 0 \\
		0 & -1 
	\epm \equiv U \delta(x) \tau_z.
	\label{Himp}
\end{align}
We solve the problem of the impurity Yu-Shiba-Rusinov (YSR) states \cite{Yu1965,Shiba1968,Rusinov1969} exactly using the T-matrix formalism described above (see also Refs.~\cite{Pientka2013,Kaladzhyan2016a,Kaladzhyan2016b}). In momentum space the unperturbed retarded Green's function is given by $\mathcal{G}_0(p,E) =\left[E+i0 - \mathcal{H}^{1D}_p \right]^{-1}$, and the corresponding real-space Green's function
\begin{align*}
{\cal{G}}_0(x,E) = \int\frac{dp}{2\pi} {\cal{G}}_0(p,E)e^{i p x}
\end{align*}
We take $\mu=0$ and we compute analytically the real-space Green's function at $x=0$ which allows us to determine the energy of the YSR states as a function of the impurity potential:
\begin{align}
{\cal{G}}_0(0,E) =
	\bpm 
		E X_0(0) & 0 \\
		0 & E X_0(0)
	\epm
\end{align}
with
\begin{align}
&X_0(0)  = - \frac{1}{\sqrt{t^2-E^2}} \frac{1}{\sqrt{\Delta^2-E^2}}.
%&X_1(0) \equiv -\int\limits_{-\pi/a}^{\pi/a}\frac{dp}{2\pi} \frac{-2t \cos pa}{4t^2 \cos^2pa+\Delta_p^2 \sin^2pa - E^2}  = 0 \quad\text{due to symmetries}\\
%&X_2(0) \equiv -\int\limits_{-\pi/a}^{\pi/a}\frac{dp}{2\pi} \frac{i \varkappa p\,e^{ipx}}{4t^2 \cos^2pa+\Delta_p^2 \sin^2pa - E^2} = 0 \quad\text{due to symmetries}
\end{align} 
The energies of the impurity bound states can be obtained by calculating the poles of the T-matrix:
\begin{align}
1 \pm U \frac{1}{\sqrt{t^2-E^2}} \frac{E}{\sqrt{\Delta^2-E^2}}  =0
\end{align}
This equation yields a pair of spurious solutions outside the gap, and a pair of YSR-like solutions inside the gap:
\begin{align}
\nonumber &E_{\pm} = \\
&\pm \sqrt{\frac{1}{2} \left[t^2 + \Delta^2 + U^2- \sqrt{\left(t^2 + \Delta^2 + U^2 \right)^2 - 4 t^2 \Delta^2}\right]}
\label{YSRenergies1D}
\end{align}
When $U \to 0$ these solutions approach the edges of the gap, i.e. $E_\pm \to \pm \Delta$ (see Fig.~\ref{fig2}), whereas when $U \to \infty$ these solutions decay as 
\begin{align}
E_{\pm} =\pm\,  \frac{\Delta}{U/t}.
\end{align}

%The bound-state energies  satisify:
%\begin{equation}
%\pm \tilde{E}_{1,2}{\Omega} \im\frac{1}{S} - \frac{1}{2\Omega} \im S  = \frac{1}{U}
%\end{equation}
%We can show that by taking the limit of $U \to \infty$,  $E=0$ is a solution for this equation. We also plot the corresponding solutions of the full equation describing the poles of the T-matrix (Eq..), and we see that it does indeed goes to zero at large U.
% In what follows we prove that what remains on the left side of the equation is identically zero:
%\beq
%\im \left[ \frac{1}{\sqrt{-1 + 2 \delta^2 +2 i \delta \sqrt{1 - \delta^2}}} + \sqrt{-1 + 2 \delta^2 +2 i \delta \sqrt{1 - \delta^2}} \right],
%\nonumber
%\eeq
%where $\quad \delta \equiv \frac{m \varkappa}{p_F} \ll 1.$

%{\bf simplify this}
%We have:
%\beq
%\sqrt{-1 + 2 \delta^2 +2 i \delta \sqrt{1 - \delta^2}} = \sqrt{\left(\delta + i\sqrt{1 - \delta^2}\right)^2} = \nonumber \\
%= \sqrt{\left(\cos\phi + i \sin\phi\right)^2} = \sqrt{e^{2i\phi}} = e^{i\phi}\quad\text{for}\quad \phi \in \left[0,\pi/2\right).
%\nonumber
%\eeq
%Therefore:
%\begin{equation}
%\im \left[ \frac{1}{e^{i\phi}} + e^{i\phi} \right] = 2\im \cos\phi = 0, \quad\text{QED}.
%\end{equation}

We can also obtain the wavefunctions for the YSR states associated with the impurity using Refs. \cite{Pientka2013,Kaladzhyan2016a,Kaladzhyan2016b}:
$$
\Psi(x) \propto {\cal{G}}_0(x,E)\cdot \tau_z \cdot \Psi(0),
$$
where $\Psi(0) = (1\;0)^\mathrm{T}$ (for $E=E_+$) and $\Psi(0) = (0\;1)^\mathrm{T}$ (for $E=E_-$) are the null-space vectors of the matrix $\mathbb{I}_2 - U\tau_z \cdot {\cal{G}}_0(0,E)$.
%We first simplify:
%\begin{align*}
%\frac{p_F^2}{2m} X_0(x) + X_1(x)\Big|_{E=0} &= -\frac{1}{2\Omega|_{E=0}} \im\left(p_F^2 \frac{e^{-p_F |x|e^{i\phi}}}{p_F e^{i\phi}}  + p_F e^{i\phi} e^{-p_F |x|e^{i\phi}} \right) = \frac{m \varkappa}{\Omega|_{E=0}} e^{-p_F |x| \cos \phi} \sin\left( p_F |x|\sin\phi \right) \\
%X_2(x)\Big|_{E=0} &= -\frac{m \varkappa}{\Omega|_{E=0}} e^{-p_F |x| \cos \phi} \sin\left( p_F |x|\sin\phi \right) \sgn x
%\end{align*}
%And we get

In the case of an infinite potential the energies of the bound states $E_\pm \to \pm 0$. In what follows we consider that $x$ can only be an integer multiple of the lattice constant $a$, i.e. $x = n a$, with $n \in \mathbb{Z}$. This is a fair restriction taking into account that we work within a lattice model. This allows us to obtain an exact form for the two zero-energy wavefunctions:
\begin{align}
\label{WFanalyt1} &\Psi_1(x) \propto \bpm 1  \\ -\sgn x \epm e^{-\frac{1}{2} \ln \left( \frac{1+\Delta/t}{1-\Delta/t}\right) |x|} \sin (\frac{\pi |x|}{2}) \\  
\label{WFanalyt2} &\Psi_2(x) \propto \bpm  -\sgn x \\ 1 \epm e^{-\frac{1}{2} \ln \left( \frac{1+\Delta/t}{1-\Delta/t}\right) |x|} \sin(\frac{\pi |x|}{2}) .
\end{align}
We note that by combining states 1 and 2 we get left and right Majorana bound states, since the factors $\frac{1 \pm \sgn x}{2}$ ensure that the WF 'lives' either only on the left side or on the right side of the impurity. The Majorana coherence length is given by $\xi = \left[ \frac{1}{2} \ln \left( \frac{1+\Delta/t}{1-\Delta/t}\right) \right]^{-1}$, and  diverges as $t/\Delta$ when $\Delta \to 0$. Such behavior is expected since the Majorana bound states become more and more delocalized when reducing the value of the superconducting order parameter. 

\begin{figure}[t!]
\begin{center}
\includegraphics[width=0.49\columnwidth]{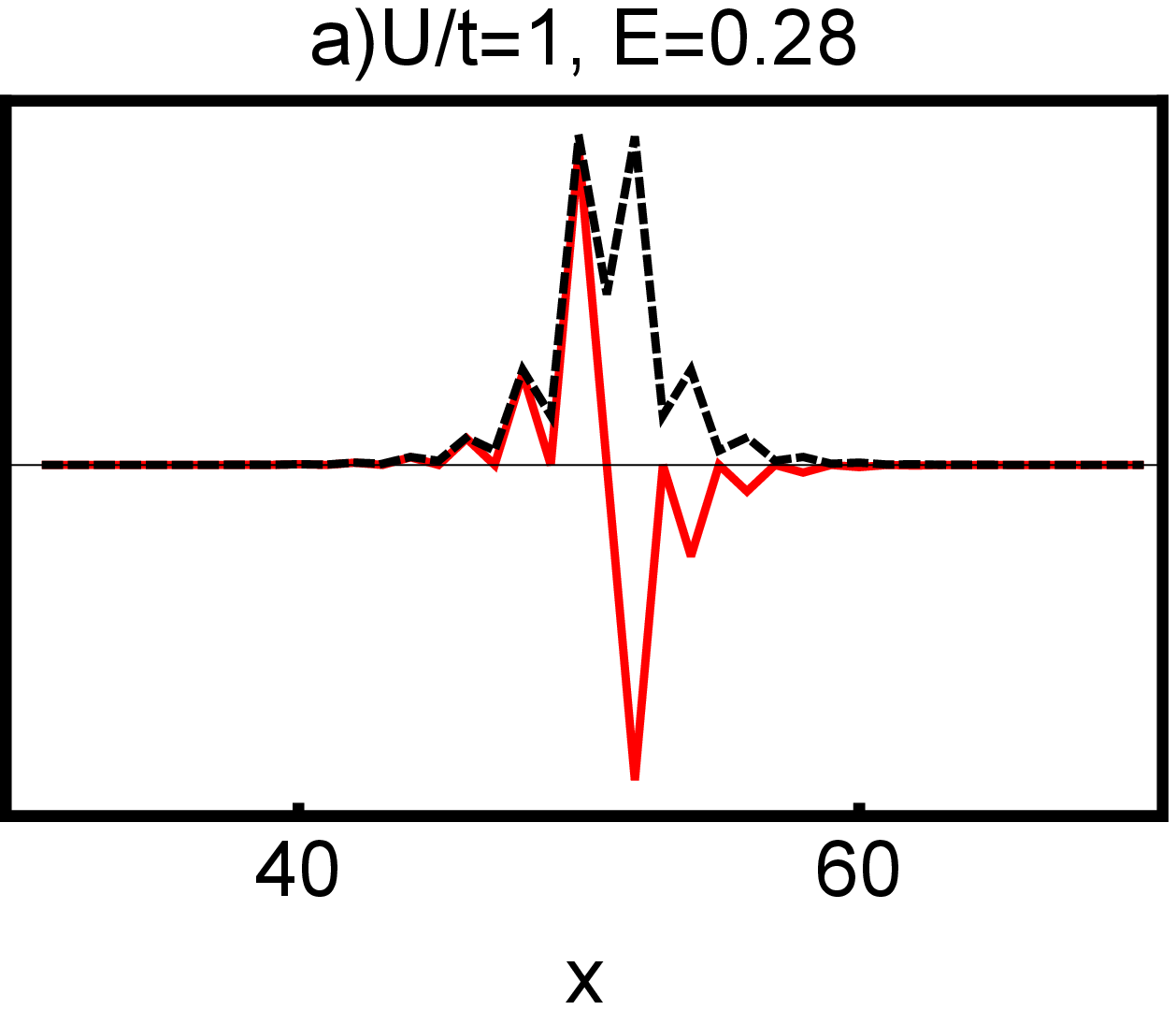}
\includegraphics[width=0.49\columnwidth]{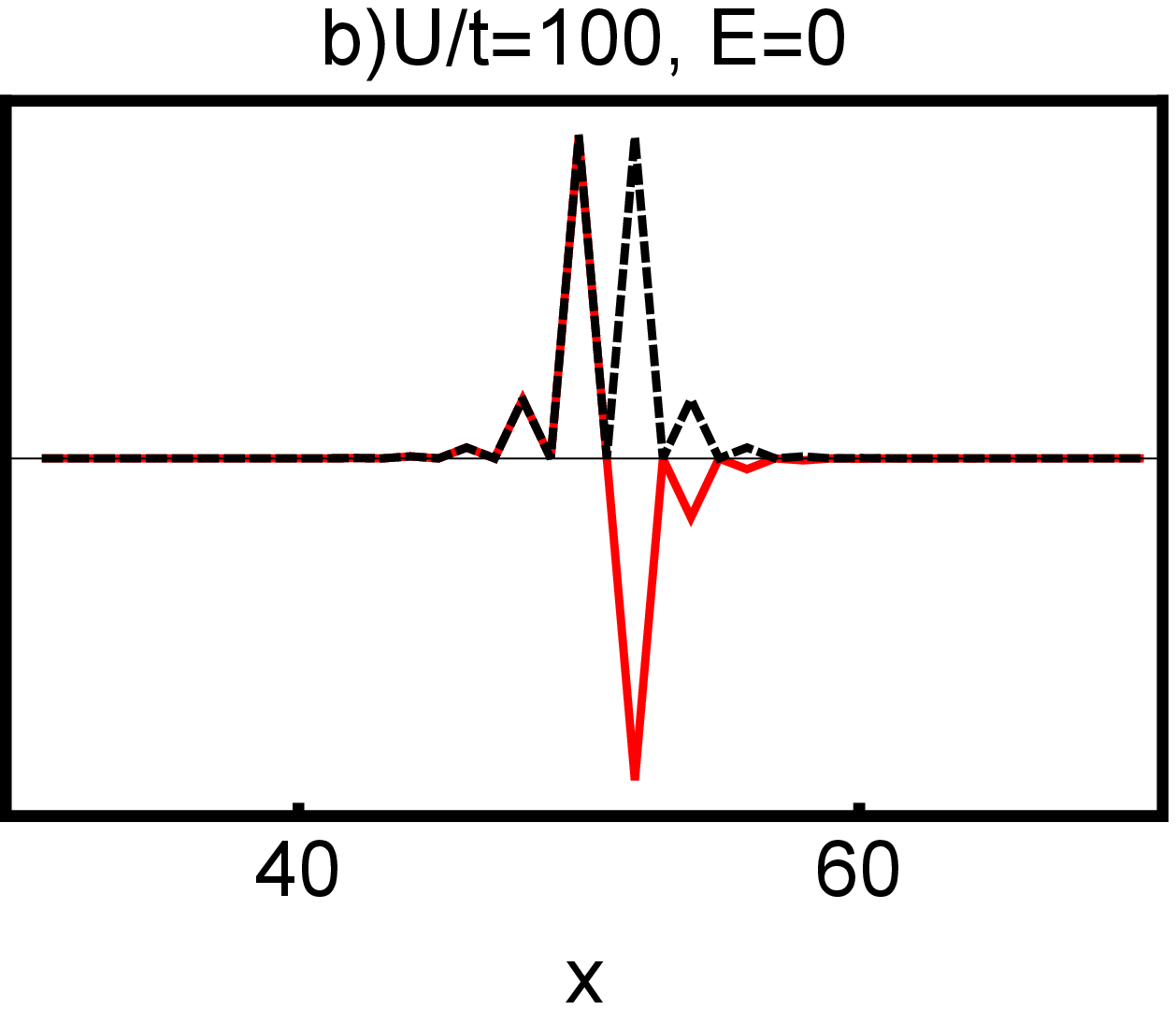}
\end{center}
\caption{The local DOS (black dashed lines) and the Majorana polarization (red solid lines) plotted as a function of position for $U/t=1$, $E/t \approx 0.28$ (left panel), and $U/t=100$, $E/t \approx 0$ (right panel). YSR states form for an impurity strength $U/t=1$, whereas for $U/t=100$ they morph into Majorana end states localized on the two sides of the potential barrier at $x=51$. We have considered a chain of 101 sites and we have set $\mu/t=0$, $\Delta/t=0.4$.}  
\label{fig3}
\end{figure}

%This is an intuitively expected result since a scalar impurity with an infinitely large potential breaks the system into two disconnected ones which in the topological phase give rise to Majorana end states. 
We confirm these findings numerically by diagonalizing a 1D Kitaev chain with an impurity using the MatQ code \cite{MatQ}, and by plotting the Majorana polarization (red line) and the local DOS (black dashed line) for the impurity-bound states (see Fig.~\ref{fig3}). The energy of the impurity-bound states goes to zero with increasing the impurity strength, for instance we get $E\approx 0.28$ and $E\approx 0$ when setting $U/t=1$ and $U/t=100$ respectively (the other parameters are  $\mu/t=0$ and $\Delta/t=0.4$). The Majorana polarization \cite{Sticlet2012,Sedlmayr2015b} differs from the density of states (DOS) for small impurities (see Fig.~\ref{fig3}a) but they become equal (up to a sign) when the impurity potential goes to infinity (Fig.~\ref{fig3}b). This indicates \cite{Sticlet2012,Sedlmayr2015b} the formation of Majorana states at the ends of the two new systems obtained by cutting the original system in two disconnected halves. Note the perfect agreement between the numerical and the analytical techniques: the one-to-one correspondence for the energies of the bound states between Figs.~\ref{fig2}, \ref{fig3} and Eq.~(\ref{YSRenergies1D}), as well as for the form of the wavefunctions in Eqs.~(\ref{WFanalyt1}) and (\ref{WFanalyt2}) versus Fig.~\ref{fig3}.\\

\textit{2D Kitaev model.} Below we turn to the case of an infinite 2D system with a  delta-like line impurity at $x=0$. We start by writing down the real-space tight-binding Hamiltonian
\begin{align}
\nonumber \mathcal{H}_{TB}^{2D} &=\negthickspace \sum\limits_{m,n}\negthickspace -\mu c^\dag_{m,n} c_{m,n} \negthickspace-\negthickspace \Big[ t \left( c^\dag_{m+1,n} c_{m,n} \negthickspace+\negthickspace c^\dag_{m,n+1} c_{m,n} \right) - \\
&\Delta \left( c_{m,n} c_{m+1,n} -i\, c_{m,n} c_{m,n+1} \right) + \mathrm{H.c.}\Big]
\end{align} 
where $\mu$ denotes the chemical potential, $t$ is the hopping parameter, and $\Delta > 0$ is the pairing amplitude. Operators $c^\dag_{m,n} (c_{m,n})$ create (annihilate) spinless fermions on the site $(m,n)$. The corresponding momentum-space lattice Hamiltonian is given by
\begin{align}
\mathcal{H}^{2D}_{\bs{p}} = 
	\bpm 
		\epsilon_{\bs{p}} & \Delta_{\bs{p}}  \\
		\Delta^*_{\bs{p}}  & -\epsilon_{\bs{p}}
	\epm,
	\label{Hp2Dlattice}
\end{align}
with $\epsilon_{\bs{p}} \equiv -\mu/2 -t \left(\cos p_x + \cos p_y \right)$, $\Delta_{\bs{p}} = i\Delta (\sin p_x + i \sin p_y)$. 

The line impurity can be described by Eq.~(\ref{Himp}). From Eq.~(\ref{Tmatrix2D}) we see that the poles of the T-matrix, which correspond to the impurity energy levels, are $p_y$-dependent and can be obtained by solving
\begin{align}
\det\left[\mathbb{I}_2 - U \tau_z \cdot \int \frac{d p_x}{2\pi} \mathcal{G}_0(p_x,p_{1y},E) \right] = 0,
\label{Tmatrixpoles2D}
\end{align}
%The most general case of $U \neq \infty$ is quite complicated, therefore, we present here only the final result in the simplest case of $U \to \infty$ while leaving the unnecessary details to the Supplementary Material \cite{SM}.
with $\mathcal{G}_0(p_x,p_{1y},E)$ being the unperturbed retarded Green's function.

At low energies we can use an approximation of the Hamiltonian in Eq.~(\ref{Hp2Dlattice}):
\begin{align}
\mathcal{H}^{2D}_{\bs{p}} \approx
	\bpm 
		\xi_p & i\varkappa (p_x + i p_y)  \\
		-i\varkappa (p_x - i p_y)  & -\xi_p
	\epm,
	\label{Hp2D}
\end{align} 
where $\xi_{\bs{p}} \equiv \frac{\bs{p}^2}{2m_0} - \frac{p_F^2}{2m_0}$ with $p_F$ denoting the Fermi momentum, $m_0$ the quasiparticle mass and $\varkappa$ the $p$-wave pairing parameter.  Such a low-energy description enables us  to obtain in the limit of $U \to \infty$ an exact analytical solution of the Eq.~(\ref{Tmatrixpoles2D}) for the poles of the T-matrix, and the result is relatively simple
\begin{align}
E_{\pm} =\pm \varkappa p_y
\end{align}
We note that when $p_y \to 0$, $E \to 0$, which is consistent with previous findings. These two solutions correspond to counterpropagating chiral Majorana modes.

\begin{figure}[t!]
\begin{center}
\includegraphics[width=0.49\columnwidth]{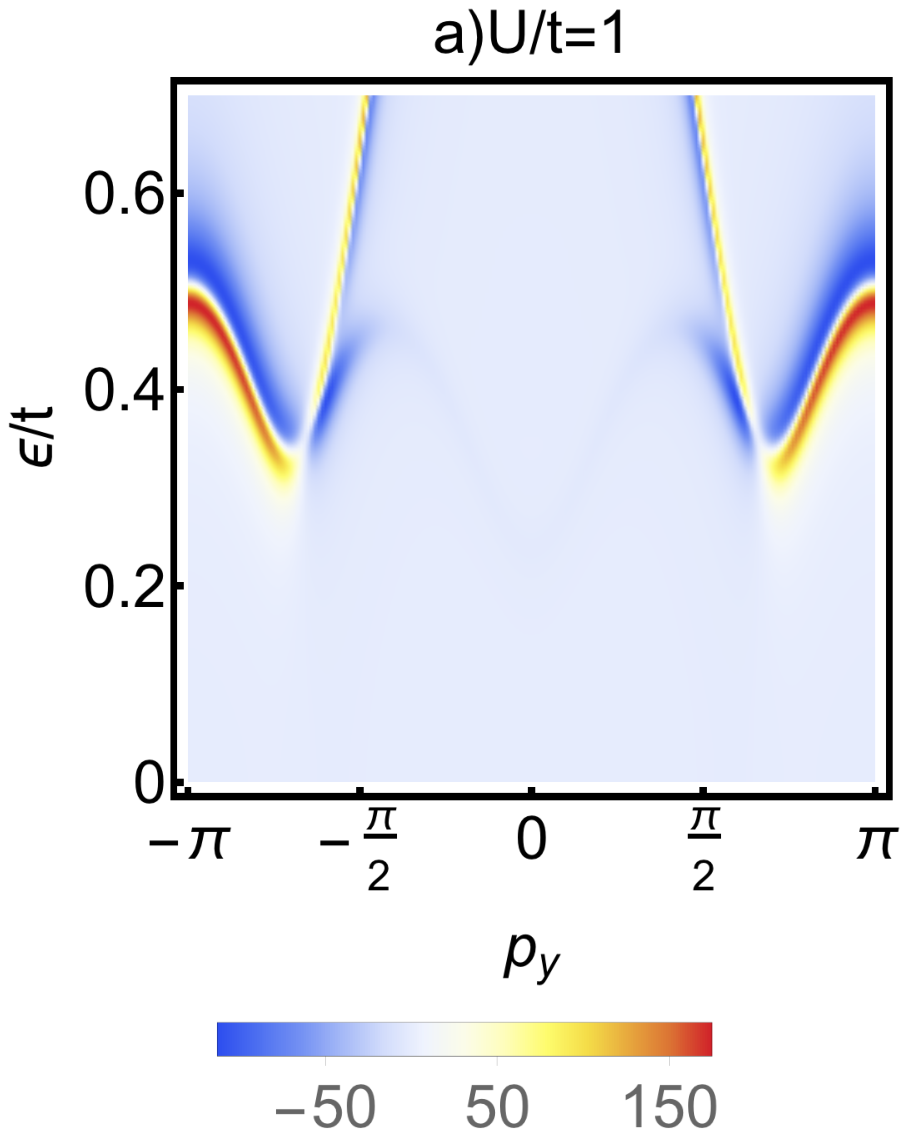}
\includegraphics[width=0.49\columnwidth]{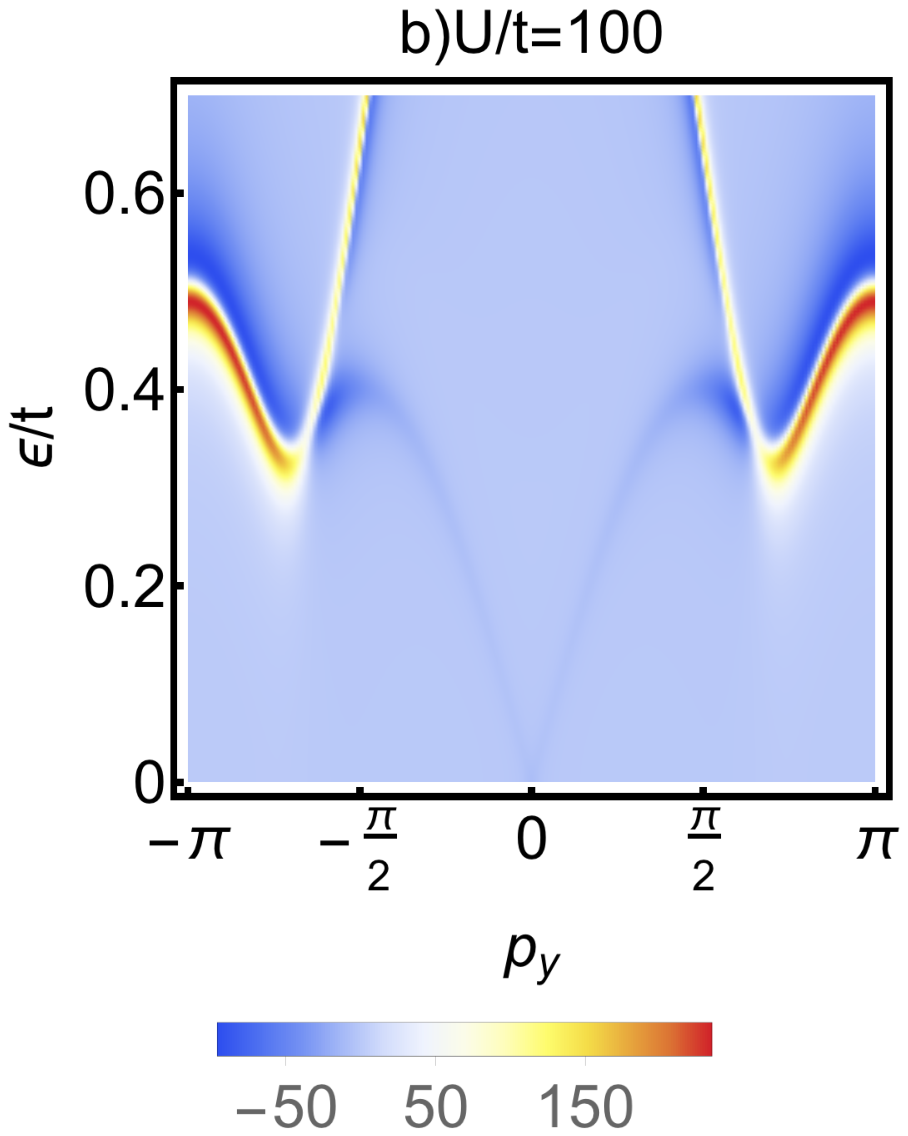}
\end{center}
\caption{On the left (right) panel we present the average spectral function of a 2D infinite system in the presence of a weak (strong) line-like impurity, plotted as a function of the momentum along the impurity $p_y$. For a weak impurity we take $U/t=1$ and we see clearly the formation of a subgap Shiba band, which for the strong impurity with $U/t=100$ morphs into chiral dispersive Majorana modes with energies $E_\pm = \pm \varkappa p_y$. We have set $\mu/t=0.5$, $\varkappa/t=0.4$, and the quasiparticle broadening $\delta/t=0.03$.}  
\label{fig4}
\end{figure}

To obtain information about the higher energy bound states we plot the the average perturbed spectral function $A(\bs{p}, E) = -\frac{1}{\pi} \im \{{\rm{Tr}} [\mathcal{G}(\bs{p}, \bs{p}, E)]\}$. The poles of the spectral function contain both the unperturbed band structure, as well as the impurity-induced bands. In order to obtain the energy dispersion of the bound states along the impurity direction we will take $p_x=0$ and plot $A(0,p_y, E)$ as a function of $p_y$ and $E$. In Fig.~\ref{fig4} we consider two values of the impurity strength $U/t=1$ and $U/t=100$. For a small impurity we note the formation of a finite-energy dispersive Shiba band (see Fig.~\ref{fig4}a), while for the very large impurity this band touches zero at $p_y=0$ (see Fig.~\ref{fig4}b), marking the separation of the system in two independent ones and the formation of chiral Majorana states. Note here that the bands above the gap correspond to the bulk unperturbed states of the system, while the subgap band is the impurity-induced band. Note also the agreement with the low-energy approximation, close to $p_y=0$ the energy dispersion of the bound states is indeed described by $E_{\pm} =\pm \varkappa p_y$ with $\varkappa/t=0.4$.

\begin{figure}%[h!]
\begin{center}
\vspace{.2in}
\includegraphics[width=0.6\columnwidth]{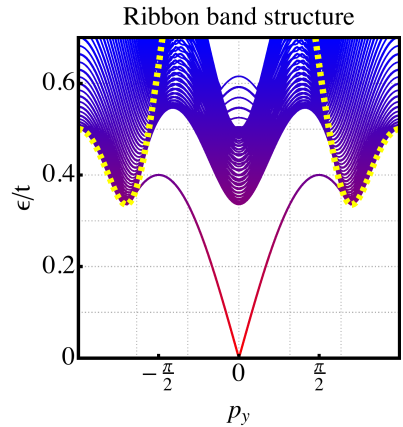}
\end{center}
\caption{The band structure of a 2D infinite ribbon obtained via numerical diagonalization of the tight-binding Hamiltonian. We note the formation of chiral dispersive Majorana modes with energies $E_\pm = \pm \varkappa p_y$, as expected theoretically for a 2D Kitaev model. Parameters are the same as in Fig.~\ref{fig4}, and  the width of the ribbon is 101 sites. The infinite system (bulk) spectrum is denoted by the dashed yellow lines.}
\label{fig5}
\end{figure}

We compare this with a fully-numerical analysis of the spectrum of a ribbon, obtained using a full tight-binding exact diagonalization via the MatQ code, and plotted in Fig.~\ref{fig5}. We note the bulk ribbon bands (denoted in blue), quantized due to the finite-size of the ribbon in the $x$ direction. For comparison we also give the infinite-system band structure superposed as dashed yellow lines. We also note the formation of the Majorana edge states crossing at $p_y=0$ (\textit{cf.} $E_{\pm} = \pm \varkappa p_y$ obtained above, denoted in red). We note the remarkable agreement between the analytical and the numerical techniques, both for the bulk, and especially for the subgap impurity states.

{\it Conclusions.} We have developed an exact formalism, which provides us with a direct manner to describe the formation of boundary modes. The technique is based on calculating the energies and the wavefunctions of the impurity-induced states in the presence of a point-like impurity (1D), line-like impurity (2D) or a plane-like impurity (3D) using the T-matrix formalism. We should point out that this formalism does not require making neither a low-energy approximation, nor any supplementary assumptions, and for the systems for which the form of the real-space Green's function can be derived analytically it does not even require a numerical integration. 

We have checked that our formalism is fully consistent with a full tight-binding numerical approach, which together with solving directly the Schr\"odinger equation with specific boundary conditions were till present the choice tools to recover the dispersion of boundary modes. We have applied our method to 1D and 2D Kitaev models to describe the formation of Majorana states, but we note that this formalism can be generalized very easily to other models supporting end, edge and surface states, for example in 3D it provides an alternative method to obtain Fermi-arc states in Weyl semimetals  \cite{unpublished2018}.

%Furthermore, we have focused on the simplest examples with point-like and line-like boundaries, however, it is worth mentioning that more complicated geometrical shapes of boundaries can be considered along similar lines, without any drastic modification of the formalism we have proposed. 

{\bf Acknowledgements}
VK would like to acknowledge the ERC Starting Grant No. 679722 and the Roland Gustafsson foundation for theoretical physics.

\bibliography{biblio_Tmatrix}

%merlin.mbs apsrev4-1.bst 2010-07-25 4.21a (PWD, AO, DPC) hacked
%Control: key (0)
%Control: author (72) initials jnrlst
%Control: editor formatted (1) identically to author
%Control: production of article title (-1) disabled
%Control: page (0) single
%Control: year (1) truncated
%Control: production of eprint (0) enabled
\begin{thebibliography}{36}%
\makeatletter
\providecommand \@ifxundefined [1]{%
 \@ifx{#1\undefined}
}%
\providecommand \@ifnum [1]{%
 \ifnum #1\expandafter \@firstoftwo
 \else \expandafter \@secondoftwo
 \fi
}%
\providecommand \@ifx [1]{%
 \ifx #1\expandafter \@firstoftwo
 \else \expandafter \@secondoftwo
 \fi
}%
\providecommand \natexlab [1]{#1}%
\providecommand \enquote  [1]{``#1''}%
\providecommand \bibnamefont  [1]{#1}%
\providecommand \bibfnamefont [1]{#1}%
\providecommand \citenamefont [1]{#1}%
\providecommand \href@noop [0]{\@secondoftwo}%
\providecommand \href [0]{\begingroup \@sanitize@url \@href}%
\providecommand \@href[1]{\@@startlink{#1}\@@href}%
\providecommand \@@href[1]{\endgroup#1\@@endlink}%
\providecommand \@sanitize@url [0]{\catcode `\\12\catcode `\$12\catcode
  `\&12\catcode `\#12\catcode `\^12\catcode `\_12\catcode `\%12\relax}%
\providecommand \@@startlink[1]{}%
\providecommand \@@endlink[0]{}%
\providecommand \url  [0]{\begingroup\@sanitize@url \@url }%
\providecommand \@url [1]{\endgroup\@href {#1}{\urlprefix }}%
\providecommand \urlprefix  [0]{URL }%
\providecommand \Eprint [0]{\href }%
\providecommand \doibase [0]{http://dx.doi.org/}%
\providecommand \selectlanguage [0]{\@gobble}%
\providecommand \bibinfo  [0]{\@secondoftwo}%
\providecommand \bibfield  [0]{\@secondoftwo}%
\providecommand \translation [1]{[#1]}%
\providecommand \BibitemOpen [0]{}%
\providecommand \bibitemStop [0]{}%
\providecommand \bibitemNoStop [0]{.\EOS\space}%
\providecommand \EOS [0]{\spacefactor3000\relax}%
\providecommand \BibitemShut  [1]{\csname bibitem#1\endcsname}%
\let\auto@bib@innerbib\@empty
%</preamble>
\bibitem [{\citenamefont {Jamtveit}\ and\ \citenamefont
  {Meakin}(1999)}]{Jamtveit1999}%
  \BibitemOpen
  \bibinfo {editor} {\bibfnamefont {B.}~\bibnamefont {Jamtveit}}\ and\ \bibinfo
  {editor} {\bibfnamefont {P.}~\bibnamefont {Meakin}},\ eds.,\ \href {\doibase
  10.1007/978-94-015-9179-9} {\emph {\bibinfo {title} {Growth, Dissolution and
  Pattern Formation in Geosystems}}}\ (\bibinfo  {publisher} {Springer
  Netherlands},\ \bibinfo {year} {1999})\BibitemShut {NoStop}%
\bibitem [{\citenamefont {Slater}\ and\ \citenamefont
  {Koster}(1954)}]{Slater1954}%
  \BibitemOpen
  \bibfield  {author} {\bibinfo {author} {\bibfnamefont {J.~C.}\ \bibnamefont
  {Slater}}\ and\ \bibinfo {author} {\bibfnamefont {G.~F.}\ \bibnamefont
  {Koster}},\ }\href {\doibase 10.1103/PhysRev.94.1498} {\bibfield  {journal}
  {\bibinfo  {journal} {Phys. Rev.}\ }\textbf {\bibinfo {volume} {94}},\
  \bibinfo {pages} {1498} (\bibinfo {year} {1954})}\BibitemShut {NoStop}%
\bibitem [{\citenamefont {Busch}\ and\ \citenamefont
  {Penson}(1987)}]{Busch1987}%
  \BibitemOpen
  \bibfield  {author} {\bibinfo {author} {\bibfnamefont {U.}~\bibnamefont
  {Busch}}\ and\ \bibinfo {author} {\bibfnamefont {K.~A.}\ \bibnamefont
  {Penson}},\ }\href {\doibase 10.1103/PhysRevB.36.9271} {\bibfield  {journal}
  {\bibinfo  {journal} {Phys. Rev. B}\ }\textbf {\bibinfo {volume} {36}},\
  \bibinfo {pages} {9271} (\bibinfo {year} {1987})}\BibitemShut {NoStop}%
\bibitem [{\citenamefont {Davison}\ and\ \citenamefont
  {St{\c{e}}{\'s}licka}(1992)}]{Davison1992}%
  \BibitemOpen
  \bibfield  {author} {\bibinfo {author} {\bibfnamefont {S.}~\bibnamefont
  {Davison}}\ and\ \bibinfo {author} {\bibfnamefont {M.}~\bibnamefont
  {St{\c{e}}{\'s}licka}},\ }\href
  {https://books.google.se/books?id=w6nvAAAAMAAJ} {\emph {\bibinfo {title}
  {Basic theory of surface states}}},\ Monographs on the physics and chemistry
  of materials\ (\bibinfo  {publisher} {Clarendon Press},\ \bibinfo {year}
  {1992})\BibitemShut {NoStop}%
\bibitem [{\citenamefont {Sancho}\ \emph {et~al.}(1984)\citenamefont {Sancho},
  \citenamefont {Sancho},\ and\ \citenamefont {Rubio}}]{Sancho1984}%
  \BibitemOpen
  \bibfield  {author} {\bibinfo {author} {\bibfnamefont {M.~P.~L.}\
  \bibnamefont {Sancho}}, \bibinfo {author} {\bibfnamefont {J.~M.~L.}\
  \bibnamefont {Sancho}}, \ and\ \bibinfo {author} {\bibfnamefont
  {J.}~\bibnamefont {Rubio}},\ }\href {\doibase 10.1088/0305-4608/14/5/016}
  {\bibfield  {journal} {\bibinfo  {journal} {Journal of Physics F: Metal
  Physics}\ }\textbf {\bibinfo {volume} {14}},\ \bibinfo {pages} {1205}
  (\bibinfo {year} {1984})}\BibitemShut {NoStop}%
\bibitem [{\citenamefont {Umerski}(1997)}]{Umerski1997}%
  \BibitemOpen
  \bibfield  {author} {\bibinfo {author} {\bibfnamefont {A.}~\bibnamefont
  {Umerski}},\ }\href {\doibase 10.1103/PhysRevB.55.5266} {\bibfield  {journal}
  {\bibinfo  {journal} {Phys. Rev. B}\ }\textbf {\bibinfo {volume} {55}},\
  \bibinfo {pages} {5266} (\bibinfo {year} {1997})}\BibitemShut {NoStop}%
\bibitem [{\citenamefont {Peng}\ \emph {et~al.}(2017)\citenamefont {Peng},
  \citenamefont {Bao},\ and\ \citenamefont {von Oppen}}]{Peng2017}%
  \BibitemOpen
  \bibfield  {author} {\bibinfo {author} {\bibfnamefont {Y.}~\bibnamefont
  {Peng}}, \bibinfo {author} {\bibfnamefont {Y.}~\bibnamefont {Bao}}, \ and\
  \bibinfo {author} {\bibfnamefont {F.}~\bibnamefont {von Oppen}},\ }\href
  {\doibase 10.1103/PhysRevB.95.235143} {\bibfield  {journal} {\bibinfo
  {journal} {Phys. Rev. B}\ }\textbf {\bibinfo {volume} {95}},\ \bibinfo
  {pages} {235143} (\bibinfo {year} {2017})}\BibitemShut {NoStop}%
\bibitem [{\citenamefont {Rhim}\ \emph {et~al.}(2018)\citenamefont {Rhim},
  \citenamefont {Bardarson},\ and\ \citenamefont {Slager}}]{Rhim2018}%
  \BibitemOpen
  \bibfield  {author} {\bibinfo {author} {\bibfnamefont {J.-W.}\ \bibnamefont
  {Rhim}}, \bibinfo {author} {\bibfnamefont {J.~H.}\ \bibnamefont {Bardarson}},
  \ and\ \bibinfo {author} {\bibfnamefont {R.-J.}\ \bibnamefont {Slager}},\
  }\href {\doibase 10.1103/PhysRevB.97.115143} {\bibfield  {journal} {\bibinfo
  {journal} {Phys. Rev. B}\ }\textbf {\bibinfo {volume} {97}},\ \bibinfo
  {pages} {115143} (\bibinfo {year} {2018})}\BibitemShut {NoStop}%
\bibitem [{\citenamefont {Balatsky}\ \emph {et~al.}(2006)\citenamefont
  {Balatsky}, \citenamefont {Vekhter},\ and\ \citenamefont
  {Zhu}}]{Balatsky2006}%
  \BibitemOpen
  \bibfield  {author} {\bibinfo {author} {\bibfnamefont {A.~V.}\ \bibnamefont
  {Balatsky}}, \bibinfo {author} {\bibfnamefont {I.}~\bibnamefont {Vekhter}}, \
  and\ \bibinfo {author} {\bibfnamefont {J.-X.}\ \bibnamefont {Zhu}},\ }\href
  {\doibase 10.1103/RevModPhys.78.373} {\bibfield  {journal} {\bibinfo
  {journal} {Rev. Mod. Phys.}\ }\textbf {\bibinfo {volume} {78}},\ \bibinfo
  {pages} {373} (\bibinfo {year} {2006})}\BibitemShut {NoStop}%
\bibitem [{\citenamefont {Kitaev}(2001)}]{Kitaev2001}%
  \BibitemOpen
  \bibfield  {author} {\bibinfo {author} {\bibfnamefont {A.~Y.}\ \bibnamefont
  {Kitaev}},\ }\href {http://stacks.iop.org/1063-7869/44/i=10S/a=S29}
  {\bibfield  {journal} {\bibinfo  {journal} {Physics-Uspekhi}\ }\textbf
  {\bibinfo {volume} {44}},\ \bibinfo {pages} {131} (\bibinfo {year}
  {2001})}\BibitemShut {NoStop}%
\bibitem [{\citenamefont {Volovik}(1999)}]{Volovik1999}%
  \BibitemOpen
  \bibfield  {author} {\bibinfo {author} {\bibfnamefont {G.~E.}\ \bibnamefont
  {Volovik}},\ }\href {\doibase 10.1134/1.568223} {\bibfield  {journal}
  {\bibinfo  {journal} {Journal of Experimental and Theoretical Physics
  Letters}\ }\textbf {\bibinfo {volume} {70}},\ \bibinfo {pages} {609}
  (\bibinfo {year} {1999})}\BibitemShut {NoStop}%
\bibitem [{\citenamefont {Read}\ and\ \citenamefont {Green}(2000)}]{Read2000}%
  \BibitemOpen
  \bibfield  {author} {\bibinfo {author} {\bibfnamefont {N.}~\bibnamefont
  {Read}}\ and\ \bibinfo {author} {\bibfnamefont {D.}~\bibnamefont {Green}},\
  }\href {\doibase 10.1103/PhysRevB.61.10267} {\bibfield  {journal} {\bibinfo
  {journal} {Phys. Rev. B}\ }\textbf {\bibinfo {volume} {61}},\ \bibinfo
  {pages} {10267} (\bibinfo {year} {2000})}\BibitemShut {NoStop}%
\bibitem [{\citenamefont {Ivanov}(2001)}]{Ivanov2001}%
  \BibitemOpen
  \bibfield  {author} {\bibinfo {author} {\bibfnamefont {D.~A.}\ \bibnamefont
  {Ivanov}},\ }\href {\doibase 10.1103/PhysRevLett.86.268} {\bibfield
  {journal} {\bibinfo  {journal} {Phys. Rev. Lett.}\ }\textbf {\bibinfo
  {volume} {86}},\ \bibinfo {pages} {268} (\bibinfo {year} {2001})}\BibitemShut
  {NoStop}%
\bibitem [{\citenamefont {Oreg}\ \emph {et~al.}(2010)\citenamefont {Oreg},
  \citenamefont {Refael},\ and\ \citenamefont {von Oppen}}]{Oreg2010}%
  \BibitemOpen
  \bibfield  {author} {\bibinfo {author} {\bibfnamefont {Y.}~\bibnamefont
  {Oreg}}, \bibinfo {author} {\bibfnamefont {G.}~\bibnamefont {Refael}}, \ and\
  \bibinfo {author} {\bibfnamefont {F.}~\bibnamefont {von Oppen}},\ }\href
  {\doibase 10.1103/PhysRevLett.105.177002} {\bibfield  {journal} {\bibinfo
  {journal} {Phys. Rev. Lett.}\ }\textbf {\bibinfo {volume} {105}},\ \bibinfo
  {pages} {177002} (\bibinfo {year} {2010})}\BibitemShut {NoStop}%
\bibitem [{\citenamefont {Lutchyn}\ \emph {et~al.}(2010)\citenamefont
  {Lutchyn}, \citenamefont {Sau},\ and\ \citenamefont
  {Das~Sarma}}]{Lutchyn2010}%
  \BibitemOpen
  \bibfield  {author} {\bibinfo {author} {\bibfnamefont {R.~M.}\ \bibnamefont
  {Lutchyn}}, \bibinfo {author} {\bibfnamefont {J.~D.}\ \bibnamefont {Sau}}, \
  and\ \bibinfo {author} {\bibfnamefont {S.}~\bibnamefont {Das~Sarma}},\ }\href
  {\doibase 10.1103/PhysRevLett.105.077001} {\bibfield  {journal} {\bibinfo
  {journal} {Phys. Rev. Lett.}\ }\textbf {\bibinfo {volume} {105}},\ \bibinfo
  {pages} {077001} (\bibinfo {year} {2010})}\BibitemShut {NoStop}%
\bibitem [{\citenamefont {Fu}\ and\ \citenamefont {Kane}(2008)}]{Fu2008}%
  \BibitemOpen
  \bibfield  {author} {\bibinfo {author} {\bibfnamefont {L.}~\bibnamefont
  {Fu}}\ and\ \bibinfo {author} {\bibfnamefont {C.~L.}\ \bibnamefont {Kane}},\
  }\href {\doibase 10.1103/PhysRevLett.100.096407} {\bibfield  {journal}
  {\bibinfo  {journal} {Phys. Rev. Lett.}\ }\textbf {\bibinfo {volume} {100}},\
  \bibinfo {pages} {096407} (\bibinfo {year} {2008})}\BibitemShut {NoStop}%
\bibitem [{\citenamefont {Qi}\ and\ \citenamefont {Zhang}(2011)}]{Qi2011}%
  \BibitemOpen
  \bibfield  {author} {\bibinfo {author} {\bibfnamefont {X.-L.}\ \bibnamefont
  {Qi}}\ and\ \bibinfo {author} {\bibfnamefont {S.-C.}\ \bibnamefont {Zhang}},\
  }\href {\doibase 10.1103/RevModPhys.83.1057} {\bibfield  {journal} {\bibinfo
  {journal} {Rev. Mod. Phys.}\ }\textbf {\bibinfo {volume} {83}},\ \bibinfo
  {pages} {1057} (\bibinfo {year} {2011})}\BibitemShut {NoStop}%
\bibitem [{\citenamefont {Yao}\ \emph {et~al.}(2009)\citenamefont {Yao},
  \citenamefont {Yang},\ and\ \citenamefont {Niu}}]{Yao2009}%
  \BibitemOpen
  \bibfield  {author} {\bibinfo {author} {\bibfnamefont {W.}~\bibnamefont
  {Yao}}, \bibinfo {author} {\bibfnamefont {S.~A.}\ \bibnamefont {Yang}}, \
  and\ \bibinfo {author} {\bibfnamefont {Q.}~\bibnamefont {Niu}},\ }\href
  {\doibase 10.1103/PhysRevLett.102.096801} {\bibfield  {journal} {\bibinfo
  {journal} {Phys. Rev. Lett.}\ }\textbf {\bibinfo {volume} {102}},\ \bibinfo
  {pages} {096801} (\bibinfo {year} {2009})}\BibitemShut {NoStop}%
\bibitem [{\citenamefont {Wan}\ \emph {et~al.}(2011)\citenamefont {Wan},
  \citenamefont {Turner}, \citenamefont {Vishwanath},\ and\ \citenamefont
  {Savrasov}}]{Wan2011}%
  \BibitemOpen
  \bibfield  {author} {\bibinfo {author} {\bibfnamefont {X.}~\bibnamefont
  {Wan}}, \bibinfo {author} {\bibfnamefont {A.~M.}\ \bibnamefont {Turner}},
  \bibinfo {author} {\bibfnamefont {A.}~\bibnamefont {Vishwanath}}, \ and\
  \bibinfo {author} {\bibfnamefont {S.~Y.}\ \bibnamefont {Savrasov}},\ }\href
  {\doibase 10.1103/PhysRevB.83.205101} {\bibfield  {journal} {\bibinfo
  {journal} {Phys. Rev. B}\ }\textbf {\bibinfo {volume} {83}},\ \bibinfo
  {pages} {205101} (\bibinfo {year} {2011})}\BibitemShut {NoStop}%
\bibitem [{\citenamefont {Okugawa}\ and\ \citenamefont
  {Murakami}(2014)}]{Okugawa2014}%
  \BibitemOpen
  \bibfield  {author} {\bibinfo {author} {\bibfnamefont {R.}~\bibnamefont
  {Okugawa}}\ and\ \bibinfo {author} {\bibfnamefont {S.}~\bibnamefont
  {Murakami}},\ }\href {\doibase 10.1103/PhysRevB.89.235315} {\bibfield
  {journal} {\bibinfo  {journal} {Phys. Rev. B}\ }\textbf {\bibinfo {volume}
  {89}},\ \bibinfo {pages} {235315} (\bibinfo {year} {2014})}\BibitemShut
  {NoStop}%
\bibitem [{\citenamefont {Pinon}\ \emph {et~al.}()\citenamefont {Pinon},
  \citenamefont {Sarkar}, \citenamefont {Kaladzhyan},\ and\ \citenamefont
  {Bena}}]{unpublished2018}%
  \BibitemOpen
  \bibfield  {author} {\bibinfo {author} {\bibfnamefont {S.}~\bibnamefont
  {Pinon}}, \bibinfo {author} {\bibfnamefont {S.}~\bibnamefont {Sarkar}},
  \bibinfo {author} {\bibfnamefont {V.}~\bibnamefont {Kaladzhyan}}, \ and\
  \bibinfo {author} {\bibfnamefont {C.}~\bibnamefont {Bena}},\ }\href@noop {}
  {\bibinfo  {journal} {in preparation}\ }\BibitemShut {NoStop}%
\bibitem [{\citenamefont {Slager}\ \emph {et~al.}(2015)\citenamefont {Slager},
  \citenamefont {Rademaker}, \citenamefont {Zaanen},\ and\ \citenamefont
  {Balents}}]{Slager2015}%
  \BibitemOpen
\bibfield  {journal} {  }\bibfield  {author} {\bibinfo {author} {\bibfnamefont
  {R.-J.}\ \bibnamefont {Slager}}, \bibinfo {author} {\bibfnamefont
  {L.}~\bibnamefont {Rademaker}}, \bibinfo {author} {\bibfnamefont
  {J.}~\bibnamefont {Zaanen}}, \ and\ \bibinfo {author} {\bibfnamefont
  {L.}~\bibnamefont {Balents}},\ }\href {\doibase 10.1103/PhysRevB.92.085126}
  {\bibfield  {journal} {\bibinfo  {journal} {Phys. Rev. B}\ }\textbf {\bibinfo
  {volume} {92}},\ \bibinfo {pages} {085126} (\bibinfo {year}
  {2015})}\BibitemShut {NoStop}%
\bibitem [{\citenamefont {Mahan}(2000)}]{Mahan2000}%
  \BibitemOpen
  \bibfield  {author} {\bibinfo {author} {\bibfnamefont {G.~D.}\ \bibnamefont
  {Mahan}},\ }\href {\doibase 10.1007/978-1-4757-5714-9} {\emph {\bibinfo
  {title} {Many-Particle Physics}}}\ (\bibinfo  {publisher} {Springer {US}},\
  \bibinfo {year} {2000})\BibitemShut {NoStop}%
\bibitem [{\citenamefont {Byers}\ \emph {et~al.}(1993)\citenamefont {Byers},
  \citenamefont {Flatt\'e},\ and\ \citenamefont {Scalapino}}]{Byers1993}%
  \BibitemOpen
  \bibfield  {author} {\bibinfo {author} {\bibfnamefont {J.~M.}\ \bibnamefont
  {Byers}}, \bibinfo {author} {\bibfnamefont {M.~E.}\ \bibnamefont {Flatt\'e}},
  \ and\ \bibinfo {author} {\bibfnamefont {D.~J.}\ \bibnamefont {Scalapino}},\
  }\href {\doibase 10.1103/PhysRevLett.71.3363} {\bibfield  {journal} {\bibinfo
   {journal} {Phys. Rev. Lett.}\ }\textbf {\bibinfo {volume} {71}},\ \bibinfo
  {pages} {3363} (\bibinfo {year} {1993})}\BibitemShut {NoStop}%
\bibitem [{\citenamefont {Ziegler}\ \emph {et~al.}(1996)\citenamefont
  {Ziegler}, \citenamefont {Poilblanc}, \citenamefont {Preuss}, \citenamefont
  {Hanke},\ and\ \citenamefont {Scalapino}}]{Ziegler1996}%
  \BibitemOpen
  \bibfield  {author} {\bibinfo {author} {\bibfnamefont {W.}~\bibnamefont
  {Ziegler}}, \bibinfo {author} {\bibfnamefont {D.}~\bibnamefont {Poilblanc}},
  \bibinfo {author} {\bibfnamefont {R.}~\bibnamefont {Preuss}}, \bibinfo
  {author} {\bibfnamefont {W.}~\bibnamefont {Hanke}}, \ and\ \bibinfo {author}
  {\bibfnamefont {D.~J.}\ \bibnamefont {Scalapino}},\ }\href {\doibase
  10.1103/PhysRevB.53.8704} {\bibfield  {journal} {\bibinfo  {journal} {Phys.
  Rev. B}\ }\textbf {\bibinfo {volume} {53}},\ \bibinfo {pages} {8704}
  (\bibinfo {year} {1996})}\BibitemShut {NoStop}%
\bibitem [{\citenamefont {Salkola}\ \emph {et~al.}(1996)\citenamefont
  {Salkola}, \citenamefont {Balatsky},\ and\ \citenamefont
  {Scalapino}}]{Salkola1996}%
  \BibitemOpen
  \bibfield  {author} {\bibinfo {author} {\bibfnamefont {M.~I.}\ \bibnamefont
  {Salkola}}, \bibinfo {author} {\bibfnamefont {A.~V.}\ \bibnamefont
  {Balatsky}}, \ and\ \bibinfo {author} {\bibfnamefont {D.~J.}\ \bibnamefont
  {Scalapino}},\ }\href {\doibase 10.1103/PhysRevLett.77.1841} {\bibfield
  {journal} {\bibinfo  {journal} {Phys. Rev. Lett.}\ }\textbf {\bibinfo
  {volume} {77}},\ \bibinfo {pages} {1841} (\bibinfo {year}
  {1996})}\BibitemShut {NoStop}%
\bibitem [{\citenamefont {Bena}(2008)}]{Bena2008}%
  \BibitemOpen
  \bibfield  {author} {\bibinfo {author} {\bibfnamefont {C.}~\bibnamefont
  {Bena}},\ }\href {\doibase 10.1103/PhysRevLett.100.076601} {\bibfield
  {journal} {\bibinfo  {journal} {Phys. Rev. Lett.}\ }\textbf {\bibinfo
  {volume} {100}},\ \bibinfo {pages} {076601} (\bibinfo {year}
  {2008})}\BibitemShut {NoStop}%
\bibitem [{\citenamefont {Yu}(1965)}]{Yu1965}%
  \BibitemOpen
  \bibfield  {author} {\bibinfo {author} {\bibfnamefont {L.}~\bibnamefont
  {Yu}},\ }\href {\doibase 10.7498/aps.21.75} {\bibfield  {journal} {\bibinfo
  {journal} {Acta Physica Sinica}\ }\textbf {\bibinfo {volume} {21}},\ \bibinfo
  {eid} {75} (\bibinfo {year} {1965})}\BibitemShut {NoStop}%
\bibitem [{\citenamefont {Shiba}(1968)}]{Shiba1968}%
  \BibitemOpen
  \bibfield  {author} {\bibinfo {author} {\bibfnamefont {H.}~\bibnamefont
  {Shiba}},\ }\href {\doibase 10.1143/PTP.40.435} {\bibfield  {journal}
  {\bibinfo  {journal} {Progress of Theoretical Physics}\ }\textbf {\bibinfo
  {volume} {40}},\ \bibinfo {pages} {435} (\bibinfo {year} {1968})}\BibitemShut
  {NoStop}%
\bibitem [{\citenamefont {{Rusinov}}(1969)}]{Rusinov1969}%
  \BibitemOpen
  \bibfield  {author} {\bibinfo {author} {\bibfnamefont {A.~I.}\ \bibnamefont
  {{Rusinov}}},\ }\href@noop {} {\bibfield  {journal} {\bibinfo  {journal}
  {Soviet Journal of Experimental and Theoretical Physics Letters}\ }\textbf
  {\bibinfo {volume} {9}},\ \bibinfo {pages} {85} (\bibinfo {year}
  {1969})}\BibitemShut {NoStop}%
\bibitem [{\citenamefont {Pientka}\ \emph {et~al.}(2013)\citenamefont
  {Pientka}, \citenamefont {Glazman},\ and\ \citenamefont {von
  Oppen}}]{Pientka2013}%
  \BibitemOpen
  \bibfield  {author} {\bibinfo {author} {\bibfnamefont {F.}~\bibnamefont
  {Pientka}}, \bibinfo {author} {\bibfnamefont {L.~I.}\ \bibnamefont
  {Glazman}}, \ and\ \bibinfo {author} {\bibfnamefont {F.}~\bibnamefont {von
  Oppen}},\ }\href {\doibase 10.1103/PhysRevB.88.155420} {\bibfield  {journal}
  {\bibinfo  {journal} {Phys. Rev. B}\ }\textbf {\bibinfo {volume} {88}},\
  \bibinfo {pages} {155420} (\bibinfo {year} {2013})}\BibitemShut {NoStop}%
\bibitem [{\citenamefont {Kaladzhyan}\ \emph
  {et~al.}(2016{\natexlab{a}})\citenamefont {Kaladzhyan}, \citenamefont
  {Bena},\ and\ \citenamefont {Simon}}]{Kaladzhyan2016a}%
  \BibitemOpen
  \bibfield  {author} {\bibinfo {author} {\bibfnamefont {V.}~\bibnamefont
  {Kaladzhyan}}, \bibinfo {author} {\bibfnamefont {C.}~\bibnamefont {Bena}}, \
  and\ \bibinfo {author} {\bibfnamefont {P.}~\bibnamefont {Simon}},\ }\href
  {\doibase 10.1103/PhysRevB.93.214514} {\bibfield  {journal} {\bibinfo
  {journal} {Phys. Rev. B}\ }\textbf {\bibinfo {volume} {93}},\ \bibinfo
  {pages} {214514} (\bibinfo {year} {2016}{\natexlab{a}})}\BibitemShut
  {NoStop}%
\bibitem [{\citenamefont {Kaladzhyan}\ \emph
  {et~al.}(2016{\natexlab{b}})\citenamefont {Kaladzhyan}, \citenamefont
  {Bena},\ and\ \citenamefont {Simon}}]{Kaladzhyan2016b}%
  \BibitemOpen
  \bibfield  {author} {\bibinfo {author} {\bibfnamefont {V.}~\bibnamefont
  {Kaladzhyan}}, \bibinfo {author} {\bibfnamefont {C.}~\bibnamefont {Bena}}, \
  and\ \bibinfo {author} {\bibfnamefont {P.}~\bibnamefont {Simon}},\ }\href
  {http://stacks.iop.org/0953-8984/28/i=48/a=485701} {\bibfield  {journal}
  {\bibinfo  {journal} {Journal of Physics: Condensed Matter}\ }\textbf
  {\bibinfo {volume} {28}},\ \bibinfo {pages} {485701} (\bibinfo {year}
  {2016}{\natexlab{b}})}\BibitemShut {NoStop}%
\bibitem [{Mat()}]{MatQ}%
  \BibitemOpen
  \href@noop {} {\emph {\bibinfo {title} {MatQ}}},\ \bibinfo {address}
  {www.icmm.csic.es/sanjose/MathQ/MathQ.html}\BibitemShut {NoStop}%
\bibitem [{\citenamefont {Sticlet}\ \emph {et~al.}(2012)\citenamefont
  {Sticlet}, \citenamefont {Bena},\ and\ \citenamefont {Simon}}]{Sticlet2012}%
  \BibitemOpen
  \bibfield  {author} {\bibinfo {author} {\bibfnamefont {D.}~\bibnamefont
  {Sticlet}}, \bibinfo {author} {\bibfnamefont {C.}~\bibnamefont {Bena}}, \
  and\ \bibinfo {author} {\bibfnamefont {P.}~\bibnamefont {Simon}},\ }\href
  {\doibase 10.1103/PhysRevLett.108.096802} {\bibfield  {journal} {\bibinfo
  {journal} {Phys. Rev. Lett.}\ }\textbf {\bibinfo {volume} {108}},\ \bibinfo
  {pages} {096802} (\bibinfo {year} {2012})}\BibitemShut {NoStop}%
\bibitem [{\citenamefont {Sedlmayr}\ and\ \citenamefont
  {Bena}(2015)}]{Sedlmayr2015b}%
  \BibitemOpen
  \bibfield  {author} {\bibinfo {author} {\bibfnamefont {N.}~\bibnamefont
  {Sedlmayr}}\ and\ \bibinfo {author} {\bibfnamefont {C.}~\bibnamefont
  {Bena}},\ }\href {\doibase 10.1103/PhysRevB.92.115115} {\bibfield  {journal}
  {\bibinfo  {journal} {Phys. Rev. B}\ }\textbf {\bibinfo {volume} {92}},\
  \bibinfo {pages} {115115} (\bibinfo {year} {2015})}\BibitemShut {NoStop}%
\end{thebibliography}%

\end{document}